\documentstyle[preprint,epsfig,aps]{revtex}
\begin{document}

\bibliographystyle{unsrt}    % for BibTeX - sorted numerical labels

\def \as {\alpha_s}
\def \gl {\tilde{g}}
\def \ms {{\overline{\mbox{MS}}}}
\def \dr {{\overline{\mbox{DR}}}}
\newcommand{\st}{\scriptstyle}
\newcommand{\sst}{\scriptscriptstyle}
\newcommand{\mco}{\multicolumn}
\newcommand{\epp}{\epsilon^{\prime}}
\newcommand{\vep}{\varepsilon}
\newcommand{\ra}{\rightarrow}
\newcommand{\ppg}{\pi^+\pi^-\gamma}
\newcommand{\vp}{{\bf p}}
\newcommand{\ko}{K^0}
\newcommand{\kb}{\bar{K^0}}
\newcommand{\al}{\alpha}
\newcommand{\ab}{\bar{\alpha}}
\def\be{\begin{equation}}
\def\ee{\end{equation}}
\def\bea{\begin{eqnarray}}
\def\eea{\end{eqnarray}}
\def\CPbar{\hbox{{\rm CP}\hskip-1.80em{/}}}%temp replacement due to no font

\def\ap#1#2#3   {{\em Ann. Phys. (NY)} {\bf#1} (#2) #3.}
\def\apj#1#2#3  {{\em Astrophys. J.} {\bf#1} (#2) #3.}
\def\apjl#1#2#3 {{\em Astrophys. J. Lett.} {\bf#1} (#2) #3.}
\def\app#1#2#3  {{\em Acta. Phys. Pol.} {\bf#1} (#2) #3.}
\def\ar#1#2#3   {{\em Ann. Rev. Nucl. Part. Sci.} {\bf#1} (#2) #3.}
\def\cpc#1#2#3  {{\em Computer Phys. Comm.} {\bf#1} (#2) #3.}
\def\err#1#2#3  {{\it Erratum} {\bf#1} (#2) #3.}
\def\ib#1#2#3   {{\it ibid.} {\bf#1} (#2) #3.}
\def\jmp#1#2#3  {{\em J. Math. Phys.} {\bf#1} (#2) #3.}
\def\ijmp#1#2#3 {{\em Int. J. Mod. Phys.} {\bf#1} (#2) #3.}
\def\jetp#1#2#3 {{\em JETP Lett.} {\bf#1} (#2) #3.}
\def\jpg#1#2#3  {{\em J. Phys. G.} {\bf#1} (#2) #3.}
\def\mpl#1#2#3  {{\em Mod. Phys. Lett.} {\bf#1} (#2) #3.}
\def\nat#1#2#3  {{\em Nature (London)} {\bf#1} (#2) #3.}
\def\nc#1#2#3   {{\em Nuovo Cim.} {\bf#1} (#2) #3.}
\def\nim#1#2#3  {{\em Nucl. Instr. Meth.} {\bf#1} (#2) #3.}
\def\np#1#2#3   {{\em Nucl. Phys.} {\bf#1} (#2) #3.}
\def\pcps#1#2#3 {{\em Proc. Cam. Phil. Soc.} {\bf#1} (#2) #3.}
\def\pl#1#2#3   {{\em Phys. Lett.} {\bf#1} (#2) #3.}
\def\prep#1#2#3 {{\em Phys. Rep.} {\bf#1} (#2) #3.}
\def\prev#1#2#3 {{\em Phys. Rev.} {\bf#1} (#2) #3.}
\def\prl#1#2#3  {{\em Phys. Rev. Lett.} {\bf#1} (#2) #3.}
\def\prs#1#2#3  {{\em Proc. Roy. Soc.} {\bf#1} (#2) #3.}
\def\ptp#1#2#3  {{\em Prog. Th. Phys.} {\bf#1} (#2) #3.}
\def\ps#1#2#3   {{\em Physica Scripta} {\bf#1} (#2) #3.}
\def\rmp#1#2#3  {{\em Rev. Mod. Phys.} {\bf#1} (#2) #3.}
\def\rpp#1#2#3  {{\em Rep. Prog. Phys.} {\bf#1} (#2) #3.}
\def\sjnp#1#2#3 {{\em Sov. J. Nucl. Phys.} {\bf#1} (#2) #3.}
\def\spj#1#2#3  {{\em Sov. Phys. JEPT} {\bf#1} (#2) #3.}
\def\spu#1#2#3  {{\em Sov. Phys.-Usp.} {\bf#1} (#2) #3.}
\def\zp#1#2#3   {{\em Zeit. Phys.} {\bf#1} (#2) #3.}

\setcounter{secnumdepth}{2} % Number sections and subsections

%%%%%%%%%%%%%%%%%%%%%%%%%%%%%%%%%%%%%%%%%%%%%%%%%%
%                                                %
%    BEGINNING OF TEXT                           %
%                                                %
%%%%%%%%%%%%%%%%%%%%%%%%%%%%%%%%%%%%%%%%%%%%%%%%%%
   
%\hfill\vbox{\hbox{\bf US-FT/11-97}\hbox{March 1997}}
\vspace{1.cm}
\begin{center}
{\Large\bf On the detection of Ultra High Energy Neutrinos  
with the Auger Observatory}\\
\vspace{1cm}
\centerline{  K.~S. Capelle$^a$, J.~W.~Cronin$^a$, G.~Parente$^b$ and
E.~Zas$^b$ }
\centerline{$^a$\it Department of Physics and Enrico Fermi Institute,}
{\it University of Chicago, Chicago, IL 60637, USA}
\centerline{$^b$\it Departamento de F\'\i sica de Part\'\i culas,}
\centerline{\it Universidade de Santiago de Compostela, E-15706 Santiago, Spain}
\vspace{1cm}

%\vspace{1cm}
%{\Large \bf DRAFT }
\end{center}
%\vspace{.5cm}

\begin{abstract} 
We show that the Auger Air Shower Array has the potential to
detect neutrinos of energies in the $10^{19}~$eV range through horizontal air
showers. Assuming some simple conservative trigger requirements we obtain the
acceptance for horizontal air showers as induced by high energy neutrinos by
two alternative methods and we then give the expected  event rates for a
variety of neutrino fluxes as predicted in different models which are used for
reference.
\end{abstract} 
\vspace{1cm} 

\newpage \section{Introduction}

High energy neutrino detection is one of the experimental challenges in
particle astrophysics for the forthcoming years because it opens a new window
to the regions of the Universe that are otherwise shielded from us by large
amounts of matter.  It is widely believed that one of the most appropriate
techniques for neutrino detection consists of detecting the \v Cerenkov light
from muons or showers produced by the neutrino interactions in underground
water or ice. This allows the instrumentation of large enough volumes to
compensate for both the low neutrino cross section and the low fluxes expected. 
There are several projects under way to build sufficiently large detectors to
measure the expected signals from a variety of sources\cite{physrep}. 

On the other hand many years ago it was suggested that deeply penetrating high
energy particles, such as muons and neutrinos,  initiate large horizontal air
showers that can be detected at ground level \cite{berez}. At large zenith
angles the electromagnetic part of ordinary air showers, initiated by cosmic
rays of hadronic (or electromagnetic) nature, is attenuated by the atmosphere
well before reaching ground level. This should allow the identification of the
showers initiated by these deeply penetrating particles.

Recently there has been a proposal to build two 3000~km$^2$ extensive air
shower arrays, one in each hemisphere, to detect cosmic rays with energy above
$10^{19}~$eV (the Pierre Auger project)\cite{Auger}. Each array consists of a
hexagonal grid of water tanks, 3.5 m diameter and 1.2 m height, combined with
an optical fluorescence detector. The tanks are separated 1.5~km from each
other and instrumented with photodetectors to detect the \v Cerenkov light
emitted as photons and charged particles from the shower front cross it. We
will show that these detectors can also play an important and complementary
role for detecting  neutrinos of energy around $10^{19}$ eV and above. 

Neutrinos of these energies cannot penetrate  the earth so any detector
searching them must look for events with zenith angles between near horizontal
and  vertical downgoing. For showers of sufficient energy the array efficiency
is high and the low target density for neutrinos provided by the atmosphere is
compensated by the huge surface area of the planned array. The observatory will
complement other neutrino detectors in construction or planning. 

In this article we show that the Pierre Auger Observatories have an  enormous
potential to detect neutrinos in the EeV range. In section II we firstly
present some of the possible EeV neutrino fluxes that have been discussed in
the literature which will be used later in the evaluation of the event rates
expected by the Auger particle arrays. In section III we discuss the acceptance
for the horizontal air showers. Firstly we obtain a rough but intuitive
analytical expression for the acceptance of a large array, which we discuss in
the context of the Pierre Auger project. In section III.A we present an
acceptance calculation for the Auger Project based on a geometrical approach.
In section III.B an alternative and conservative calculation based on Monte
Carlo simulation is presented.  In section IV we estimate the horizontal
shower rates expected for the neutrino fluxes addressed in section II, and
section V is reserved for the conclusions.  

We do not attempt to discuss    event reconstruction or backgrounds which will
be addressed elsewhere. 

\section{EeV Neutrino fluxes}

Neutrinos of EeV energies are likely to be directly produced together with
cosmic rays of the highest energies whose origin is still a matter of
speculation.  Several possible sources of neutrinos in the EeV region have been
discussed in this context. Moreover, provided that the highest energy cosmic
rays are extragalactic in origin, as it is currently believed, EeV neutrinos
have to be produced in their interactions with cosmic microwave photons. We
will use some of the neutrino flux predictions reported in the literature as
reference calculations to evaluate the event rates that could be expected in
each Auger Observatory. In this section we briefly motivate these fluxes which
are shown in Fig.~1 compared with the expected atmospheric neutrino flux. 

Active Galactic Nuclei (AGN), the most energetic objects known, emit most of
their luminosity in gamma rays \cite{EGRET} and may be the source of the
highest energy cosmic rays \cite{biermann}. Multiwavelength observation of
these sources leads us to believe they are powered by large accreting black
holes, where shock fronts provide particle acceleration. If, besides electrons,
protons are also accelerated, as some models suggest, photoproduction of pions
with the ambient light plays a crucial role (see for example Ref.
\cite{stecker95} and references therein). The energetic gamma
rays observed come from the decays of $\pi^0$'s and high energy neutrinos from
the decays of charged pions become a signature of these models. First models
for neutrino production in AGN assumed shock acceleration in the AGN cores and
predicted relatively flat fluxes up to energies of about $10^{15}~$eV. For our
event rate calculation we select the prediction of Ref. \cite{protheroe},
labelled AGN-92C, which is quite similar to that in Ref. \cite{stecker95}
(AGN-95C).

There is however recent evidence that  the GeV to TeV gamma ray emission
observed from AGN corresponds to the blazar class \cite{Mattox}. In a unified
AGN description, the blazar class is identified as AGN  with jets of
ultrarelativistic particles streaming out of their  cores and pointing towards
us.  Most recent models for the proton blazars site the acceleration in the
jets themselves. Photoproduced neutrinos are Lorentz boosted to energies well
above those predicted in the models of acceleration in the AGN cores. We use
the  prediction of Ref.~\cite{mannheim95} (labelled AGN-95J) which illustrates
that the emitted neutrinos may extend well into the EeV region in agreement
with Ref. \cite{protheroe96}. 

We also consider the more speculative and uncertain models where the highest
energy cosmic rays are produced by the decays of topological defects. These
objects emit massive X particles, predicted in Grand Unified theories, which
decay and fragment into Standard Model particles.   There are a fair number of
neutrino flux predictions from these models depending on the different time
evolution of the effective injection rate of  X particles per unit volume
(usually denoted with a parameter $p$), the masses of the X particles
themselves and other unknown parameters.  Some models have already been
constrained by bounds on horizontal showers induced by neutrinos \cite{blanco}
and also by measurements of the 100 MeV  diffuse gamma ray background
\cite{protstan}.  We select two fluxes illustrating the range of predictions in
such models. We use the prediction with $p=1.5$ in Ref. \cite{battarch} 
(labelled TD-92) and that given in Ref. \cite{sigl} which is labelled as TD-96. 
TD-92 is the lowest neutrino prediction given in Ref.~ \cite{battarch} and is 
not severely affected by experimental constraints. 

Lastly we consider two predictions for the high energy neutrinos produced in
the interactions of cosmic rays with the cosmic microwave background: CMB-91
\cite{nuGZK} and CMB-93 corresponding to model 4 in Ref. \cite{nuGZKYosh}. They
illustrate the variations that can be associated to various uncertainties
intrinsic to these calculations. These neutrinos are a direct result of the
Greisen-Zatsepin-Kuz'min cutoff and must be produced if the highest energy
cosmic rays are of extragalactic origin. Unfortunately it is possible that the 
flux levels are too low to be observed either with the Pierre Auger Observatory
or with other neutrino detectors in planning or construction. 

When the electron neutrino flux prediction is not explicitly available in any
of the models used, we approximate it to be a factor of two below the muon
neutrino flux as expected by naive channel counting in pion production and
decay.   

\section{Acceptance for Neutrino Showers}

Neutrinos produce showers in most interactions with the atmosphere which are of
different nature depending on the process in consideration. We consider both
deep inelastic charged and neutral current interactions which always produce
hadronic type showers. In the case of charged current electron neutrino
interactions the emerging electron contributes in addition a pure
electromagnetic shower carrying a large fraction of the incoming particle
energy. We will ignore the resonant cross section because it is only
significant near the peak of the cross section which occurs at an incoming
neutrino energy of 6.4~PeV, well below the region of high efficiency for the
Pierre Auger project.

%  \begin{figure}[hbt]
%  \centering 
%  \vspace*{-2.5cm}
%  \mbox{\epsfig{figure=fig1aspen96.ps,width=12.0cm}} 
%  \vspace*{-3cm}   
%  \fcaption{Neutrino flux predictions in the EeV range.}
%  \label{fig:radk}
%  \end{figure}

For a neutrino flux $d\Phi_{\nu}/dE_{\nu}$ interacting through a process with
differential cross section $d\sigma/dy$, where $y$ is the fraction of the
incident particle energy transferred to the target, the event rate for
horizontal showers can be obtained by a simple convolution:
\begin{eqnarray}
\Phi_{sh} [E_{sh}>E_{th}] = N_a \rho_{air}  \int_{E_{th}}^{\infty} dE_{sh}
\int_{0}^{1} dy \, \frac{d \Phi_{\nu}}  {dE_{\nu}} (E_{\nu}) \, \frac{d
\sigma}{dy} (E_{\nu},y) \, {\cal A}  (y, E_{\nu})  
%\\ \nonumber \label{eq:13}
\end{eqnarray}  
where $N_a$ is Avogadro's number and $\rho_{air}$ is the air
density. The energy integral corresponds to the shower energy $E_{sh}$ which is
related to the primary neutrino energy $E_{\nu}$ in a different way depending
on the interaction being considered.  $\cal{A}$ is a geometric acceptance, a
function of shower energy, which corresponds to the volume and solid angle
integrals  for different shower positions and orientations with respect to the
array. The function is different for showers induced by charged current
electron neutrino interactions from those arising in neutral current or muon
neutrino interactions. This is  because hadronic and electromagnetic showers
have differences in the particle distributions functions, particularly for
muons.

For showers of sufficient energy, incident with zenith angle $\theta_{zenith}$,
the effective volume for horizontal shower detection is given by  $S \cos
\theta_{zenith}$, where $S$ is the surface area covered by the array, 
multiplied by the range of allowed positions for the first interaction point
along the incident direction.  We can estimate this range approximating shower
maximum as a disk of some effective radius $r$. This radius should
approximately be the maximum distance to the shower axis at which the particle
density is high enough to trigger the detector tanks (see below). As the first
interaction point is moved along the incident direction the intersection of
this disk with the detector plane is an ellipse with major axis $q=2 r/sin
\alpha$ with $\alpha=90^o-\theta_{zenith}$. The projection of this axis onto
shower direction ($q~cos \alpha$) is the wanted range (see Fig. 2).  

We estimate $\cal{A}$ integrating this volume over the possible solid angle
orientations of the shower, $d \Omega = 2 \pi d (sin \alpha)$, and restrict the
integration to horizontal showers i.e. $0^o < \alpha < \alpha_{max} \simeq
15^o$. Approximating $cos \alpha \simeq 1$ we obtain a simple analytic
expression in terms of $r$. 
\begin{eqnarray}  {\cal{A}} = S \times 2 \pi \, 
\left(   \int_{\sin \alpha_1}^{\sin \alpha_{max}} \,  d(\sin \alpha ) \sin
\alpha  \frac{2r}{\sin \alpha } + \int_{0}^{\sin \alpha_1} \,  d(\sin \alpha )
\sin \alpha~ \widehat{W} \right)
\nonumber \\  
\mbox{ }  = S \times 2 \pi \,  r (2 \sin \alpha_{max} - \sin \alpha_1) 
\label{eq:1} 
\end{eqnarray}   
For
$\alpha < \alpha_1=\sin^{-1}(2r/\widehat{W})$ the ellipse axis $q$  exceeds
$\widehat{W}$, the average size of the array~\footnote{For $S=3000~km^2$ the
''diameter'' of the Pierre Auger array is approximately $D = 65~km$ and
$\widehat{W} \simeq 0.70 D \sim 45~km$.}.  The last term in the brackets of
Eq.~(2) represents a small correction obtained when restricting the range of
the interaction point to be below the longitudinal size of the array. Most
importantly the acceptance is seen to depend on the array surface area and the
shower radius. The estimate makes the assumption that showers with shower
maximum intercepting the array are detected with close to $100 \%$ efficiency.
The effective radius has to be chosen to match this requirement which will only
hold for energies above a given threshold. It is conservative in the sense that
it ignores the fact that showers can trigger the detector without having shower
maximum intercepting the array, or even without the shower axis going through
the array (i.e. completely horizontal showers). 

It is easy to see that the acceptance for the Auger particle arrays is
comparable to other neutrino detectors in planning \cite{Venice}. Since design
requirements have led to a tank size that allows near $100\%$ efficiency for
vertical showers of energy above $10^{19}$~eV \cite{Auger}, the expected signal
in the tanks should be large enough to trigger up to distances from shower axis
of order the separation between the tanks, $1.5~$km. It is thus conservative to
expect similar efficiencies for horizontal showers above $10^{19}$~eV as they
should not differ that much from vertical ones at detector level and to
approximate $r\sim 1.5~$km. The water \v Cerenkov technique
is particularly well
suited for horizontal showers. The transverse separation between detectors will
be substantially reduced for near horizontal showers and the extra depth of the
tanks in the horizontal mode should enhance the signal from the muons in the
shower, compensating the reduction in transverse area of the tank to the
incident direction. 

Taking $r=1.5$~km ($\alpha_1 = 4^0$)  we obtain an estimate for the Auger array
acceptance of showers of energy above $\sim 10^{19}$~eV,  ${\cal{A}} =
13000~km^3~sr$ which when multiplied by an air density $\rho_{air} \simeq
10^{-3}~g~cm^{-3}$ gives $1.3~10^7~kT~sr$.  Underground high energy neutrino
detectors in planning aim towards an active volume in the range of $1~km^3$ of
water \cite{1km3}. In models where most of the events are due to neutrinos 
well above the PeV region, the Earth will be opaque to them. The corresponding
acceptance for contained events is of order $6~10^6~kT~sr$\footnote{However for
muon neutrinos the long range of the muon increases the acceptance for non
contained charged current interactions.}, illustrating how the Pierre Auger
project may come into play. 

The above estimate of the acceptance does not display the energy dependence but
it illustrates how it depends on the effective shower radius, which in turn
must depend on energy. As it happens for vertical showers, the acceptance must
show a kind of threshold behavior, since sufficiently low energy showers cannot
be detected with such a sparse array. The dependence of  the acceptance on
shower energy has been introduced in two alternative ways. In a more
geometrical approach we have followed on the above line of thinking 
considering in more detail the shower depth development and the lateral
distribution functions for electrons and positrons. As an alternative we have
simulated showers at different positions with respect to the detector in a
Monte Carlo approach.
 
\subsection{Geometrical Approach}

The calculation treats electromagnetic and hadronic cascades independently 
using the conventional NKG charged particle lateral distribution functions  
\cite{greisen} normalized with the Gaisser \cite{gaisser} (Greisen
\cite{greisen})  parametrization for the total number of electrons and
positrons in hadronic (electromagnetic) cascades. 

We assume that triggering requirements for each tank can be specified as fixed
numbers for the electron density in the electromagnetic and hadronic showers:
$\rho_e^{th}$. The relevant quantity for the tank signal is the \v Cerenkov
light, proportional to the charged particle tracklength, which is in turn
approximately proportional to energy deposition. Muons travel through the tank
depositing about $2$ MeV/(g~cm$^{-2}$) while most electrons and photons are
stopped in the tanks, depositing all their energy. We use $\rho_e^{th}$ because
there are convenient parameterizations in the literature for this quantity. The
relation between $\rho_e^{th}$ and energy deposition is however dependent on
the shower position and can only be done in an approximate manner. 

The above assumption naturally defines the active region of a shower, the
volume within which the electron density exceeds the given threshold
$\rho_e^{th}$. The region is bounded by a cigar shaped contour plot of the
three dimensional electron distribution function at  $\rho_e^{th}$. For the
solid angle-volume acceptance integration, we consider the number of tanks
contained in the intersection of these active volumes with the detector plane
as the shower directions and first interaction point are varied. The
electron-positron density at these tanks is by definition above threshold.  At
typical large distances (of order a kilometer) to the shower axis, we estimate
that an electron  density in the range 0.6-1.2~m$^{-2}$ (0.3-0.5~m$^{-2}$) for
horizontal  electromagnetic (hadronic) showers is equivalent to an energy
deposition of 500~MeV in a tank (corresponding to about two vertical muons).
This is considered sufficient for shower detection \cite{Auger}. 

Approximating the intersections of these active regions as rectangles allows a
complex but analytical solution to the problem if the only orientations and
positions of the rectangles considered in the integration are those which
contain at least $n$ tanks in the same row (see Fig.~3). As a result we obtain
the acceptance curves as a function of shower energy for each shower
distribution function which only depend on the parameter $\rho_e^{th}$.  The
important advantage of this calculational method against the obvious
alternative which involves Monte Carlo simulation (addressed below) is
computing speed. The approach converts in straightforward the otherwise lengthy
evaluation of the effect of changing the input parameters in the calculation
such as trigger conditions, tank thresholds, parameterizations of shower
distribution functions and even array spacing. 

In Fig.~4 we show the acceptance results for a trigger of at least $n=3$
aligned tanks, having the conservative electron density of 1.2~m$^{-2}$
(0.5~m$^{-2}$) for electromagnetic (hadronic) showers and  considering only
showers with zenith angle higher than 75 degrees. For each shower type there
are two curves, one considers showers with axis falling within the array and
the second also includes an approximate calculation of the triggering showers
with axis not going through the array. It is important to note that these
showers make most of the contribution to the acceptance in the high energy
limit (corresponding to huge showers flying parallel and on top of the array). 

Besides using $\rho_e^{th}$ as the parameter for tank threshold, the geometrical
algorithm described above has a potentially more important drawback. By using
parameterizations for air showers, it ignores the effect of the ground on
shower development. The simplification can be thought to grossly overestimate
the acceptance. Using EGS4 with  constant density air, we have simulated the
effect of the ground for very inclined electromagnetic showers of energies up
to $10^{15}~eV$. In spite of the reduction in the lateral distribution observed
after the core hits the ground, there is only a very small reduction in the
acceptance assuming the results can be extrapolated to higher energies. In any
case we have decided to contrast the acceptance results against a totally
different calculational approach, which we chose to remedy these drawbacks in a
conservative manner. That way we hope to bracket a prediction for the
horizontal shower acceptance of the Auger project.  

\subsection{Monte Carlo approach}

We have performed a totally independent and conservative calculation with a
hybrid Monte Carlo and parametrization method. The approach uses  the simulated
curves of energy deposition versus distance from shower axis in  the Pierre
Auger water tanks for a $10^{19}~$eV shower at different depths. These results
are then simply scaled with shower energy. We have restricted the range of
depths for the shower to the interval [500,1600]~g~cm$^{-2}$ and we   only
consider showers with the shower axis falling within the array. In our effort
to be conservative we completely neglect the contributions from particles after
the shower core hits the ground. We have then generated horizontal events by
allowing the starting point and direction ($75^o< \theta_{zenith} < 90^o$) of
the shower to vary in a random way, only restricted by the array size and a
maximum height of 3 km. 
% 
%\begin{eqnarray} %75^o< \theta_{zenith} < 90^o \;\;\;\; \;\;\; 
%                        \mbox{(near horizontal)} 
%\nonumber \\
%0^o < \phi (degrees) < 60^o  \;\;\;\;\;\;\; \mbox{(due to 3-fold symmetry)} 
%\\ 
%0<z<3 \;\;,\;\;\; 0<x<1.5 \;\;,\;\;\; 0<y<1.5 \sin 60^o \;\;\;\; 
%\mbox{(in Km)} 
%\nonumber 
%\end{eqnarray}  
%

In the simulation of the acceptance, events are detected when the shower
deposits more than 500 MeV of energy in each of four or more tanks. We have
used an average parameterization assuming half of the energy induces a hadronic
shower and the other half an electromagnetic shower. Besides correcting
drawbacks in the geometrical approach, the method has the advantage of allowing
the study of the different patterns of hit tanks on an event by event
basis\cite{capelle}. The results are shown as points in Fig. 4 for some
discrete energies. Given the fundamental differences between the two
approaches, it is not possible to test one against the other. Nevertheless it
is fair to remark that the agreement is rather good and that the differences of
results can be interpreted on the basis of the different inputs intrinsic to
each method. We thus take the Monte Carlo approach as a lower bound of the
acceptance and the difference between the two approaches can be considered as
an indication of the degree of uncertainty. 

For the detection of the horizontal showers included in the acceptance
calculations described here, only directional reconstruction and an ability to
separate the electromagnetic part of the shower from the muonic component
should be sufficient in principle. Nevertheless if the long range of the muons
produced in the horizontal showers of hadronic nature is also taken into
account, it can be argued that the acceptance may exceed that actually
calculated in this work. The muon component effectively enhances the length of
the active region of the shower, what must increase the acceptance for deeply
penetrating particles. The identification of such showers would become however
more complicated, because they look more like those induced by cosmic rays at
very large zenith angles. 

\section{Event Rates}

To calculate event rates we use the deep inelastic charged and neutral current
cross-section from two sets of structure functions  MRS(R1) and GRV ($\ms$
renormalization scheme) parton distributions  \cite{DIS}. With increasing
energy, neutrinos interact with partons carrying a lower fraction $x$ of
nucleon momentum which extends beyond the kinematical limits of the parton
density parameterizations. For the first set we extrapolate to low $x$ beyond
validity of the parameterization, using the slope of $xq(x)$, at the lowest $x$
permitted (in this case $x = 10^{-5}$), where $q(x)$ is the standard parton
distribution. This is consistent with extrapolations based on the leading log
approximation \cite{parenteICRC}. The second set, GRV, can be cautiously used
on its own for low $x$.
 
Table~1 displays the event rates for different neutrinos fluxes, calculated
with Eq.~(1) for the each of the two cross sections. The first entry 
corresponds to the result of our conservative approach based on the simulation
and the higher one to the acceptance results calculated in the geometrical
approach including showers with axis not intercepting the array. Recent models
of proton acceleration in AGN jets (AGN-95J) and some models of decays of
topological defects (such as TD-92) predict neutrino fluxes giving measurable
rates in the most conservative assumptions.  For neutrinos from AGN cores the
majority of the detected showers lie in the  $10^{15}-10^{17}$~eV region. This
corresponds to the threshold region for the acceptance curves where the
efficiency is low and are very sensitive to trigger conditions. For these
neutrino fluxes, a $1~km^3$ conventional underground neutrino detector is
expected to give more events. The detection of neutrinos from interactions of
cosmic rays with the cosmic ray background is unfortunately only marginal for
the most optimistic predictions.  It should be mentioned that conventional muon
underground detectors in planning will have similar difficulties, if not
higher, in the detection of these neutrinos.  
%may actually be impossible\cite{nuGZKYosh}**??**. 

\section{Conclusions}

The Pierre Auger project can be made sensitive to ultra high energy neutrino
fluxes through horizontal showers if an appropriate trigger is implemented. Its
acceptance for detecting contained neutrino events of energy above $E_{\nu}
\sim 10^{19}~eV$ can be very large and comparable to neutrino telescopes in 
planning.  The highest efficiency for horizontal shower detection with the
Pierre Auger Project is expected at  energies about $10^{19}~eV$, what makes it
in principle a  tool for the search of the  neutrinos from interactions of
cosmic rays with the cosmic microwave background. The Auger array can detect
the very high energy neutrinos from the decays of topological defects in some
predictions.   In any case the plethora of topological defect models will be
further constrained by the neutrino detection capabilities of the Auger 
observatories. Most recent predictions for neutrinos produced by protons
accelerated in the jets of AGN have also prospects to be detected as horizontal
shower events in the Auger particle arrays. 

In any case the target mass of the Auger detector is sufficiently large to
serve as a significant explorating tool for ultra high energy neutrinos, what
ever their source.

{\bf Acknowledgements}

We thank J. Alvarez Mu\~niz, J.J. Blanco Pillado, P. Billoir, F. Halzen, A.
Letessier-Selvon, K.~Mannheim, J. Mathews, D. Nitz, R.~Protheroe, C. Pryke,
G.~Sigl, R. V\'azquez and A.A.~Watson for a number of discussions and
suggestions. This work was supported in part by CICYT (AEN96-1673) and by Xunta
de Galicia (XUGA-20604A96). K.S.C. was supported by NSF grant PHY 9506751
``Research Experiences for Undergraduates''.

\newpage

\begin{table}[hbt] 
\label{ratenu} 
% \begin{center}               %  <-- not needed--displaymath centers
\begin{displaymath} 
\begin{array}{||c||c|c|c||}
\hline \hline \nu~source  & {\rm MRS(R1)} & {\rm GRV 95} & {\rm Energy (GeV)}\\
\hline {AGN-92C} \cite{protheroe} &  2.5/0.4  &  2/0.3  & 10^{6} < E < 10^{8}\\
{AGN-95C} \cite{stecker95} &  1.5/0.2  &  1.2 /0.2 & 10^{6} < E < 10^{8}  \\
{AGN-95J} \cite{mannheim95}&  7/2.1    &  5.5 /1.6 & 10^{8} < E < 10^{10} \\
{CMB-91} \cite{nuGZK}      & 0.5/0.2   &  0.4/0.1  & 10^{10} < E < 10^{11} \\
{CMB-93} \cite{nuGZKYosh}  &  0.2/0.06 & 0.13/0.04 & 10^{8} < E < 10^{12} \\
{TD-92} \cite{battarch}    &  12/4     &   7/2.4   & 10^{9} < E < 10^{13} \\
{TD-96} \cite{sigl}        & 1.4/0.5   & 0.9/ 0.3  & 10^{10} < E < 10^{12} \\
%  & & & & & & \\ 
\hline 
%\hline 
\end{array} 
\end{displaymath}
%\end{center} 
\caption{Yearly neutrino event rates for different diffuse fluxes
and cross sections (see text). The first entry corresponds to the upper set of
curves in Fig.~2 (all showers) and the more conservative second entry is the
result of the simulation. The last column gives the energy range of the bulk of
the events. Events correspond to a single array of 3000~km$^2$.} 
\end{table}

\vspace{2.cm}

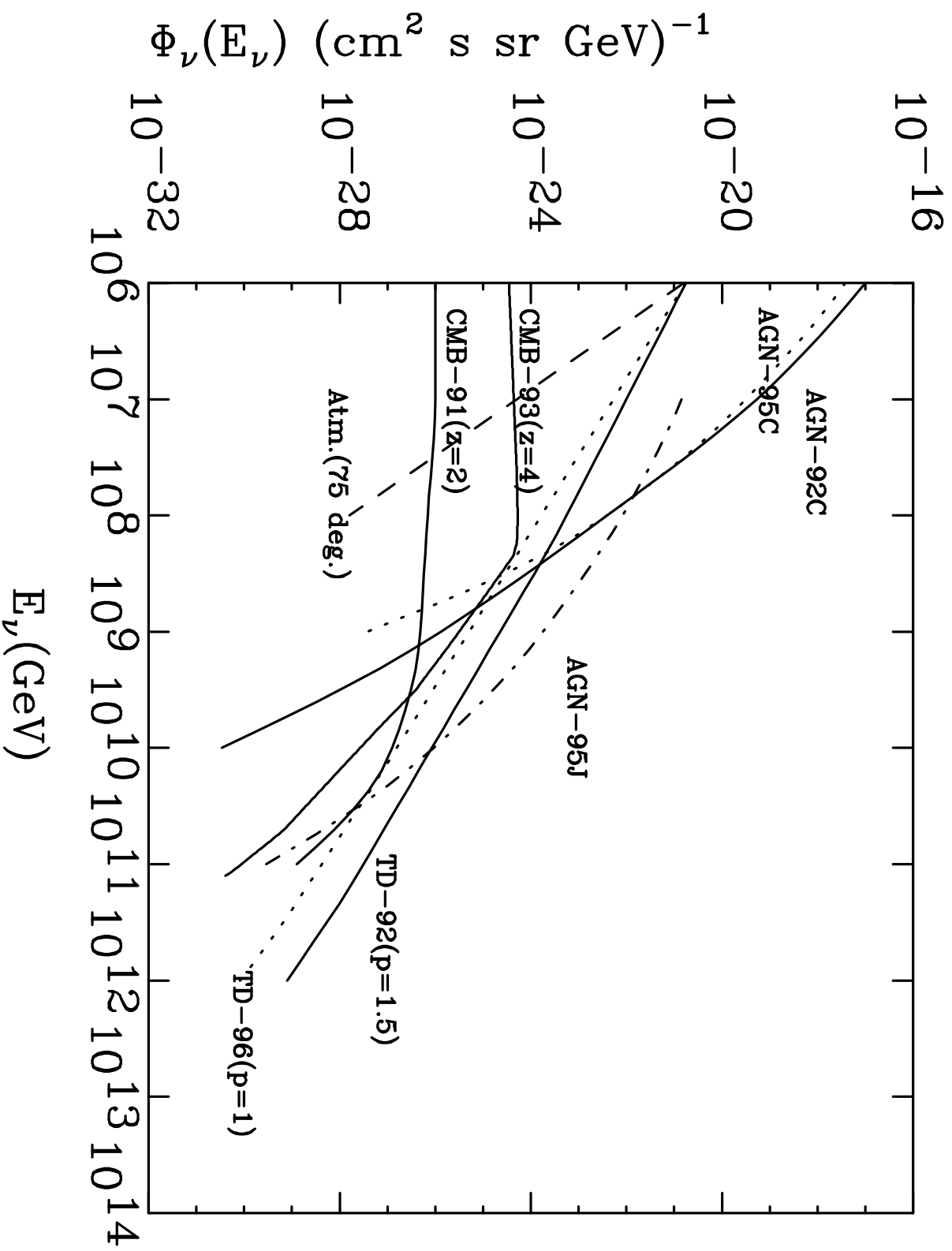
\noindent{Figure 1: Neutrino flux predictions in the EeV range as labeled in
the text.}

\vspace{1.cm}

\noindent{Figure 2: Schematic picture of a horizontal shower, illustrating the
disk of radius $r$ at shower maximum and the cylinder it spans as the first
interaction point is shifted along the incident direction. The intersection of
the cylinder is an ellipse of major axis $q$ (see text).}

\vspace{1.cm}

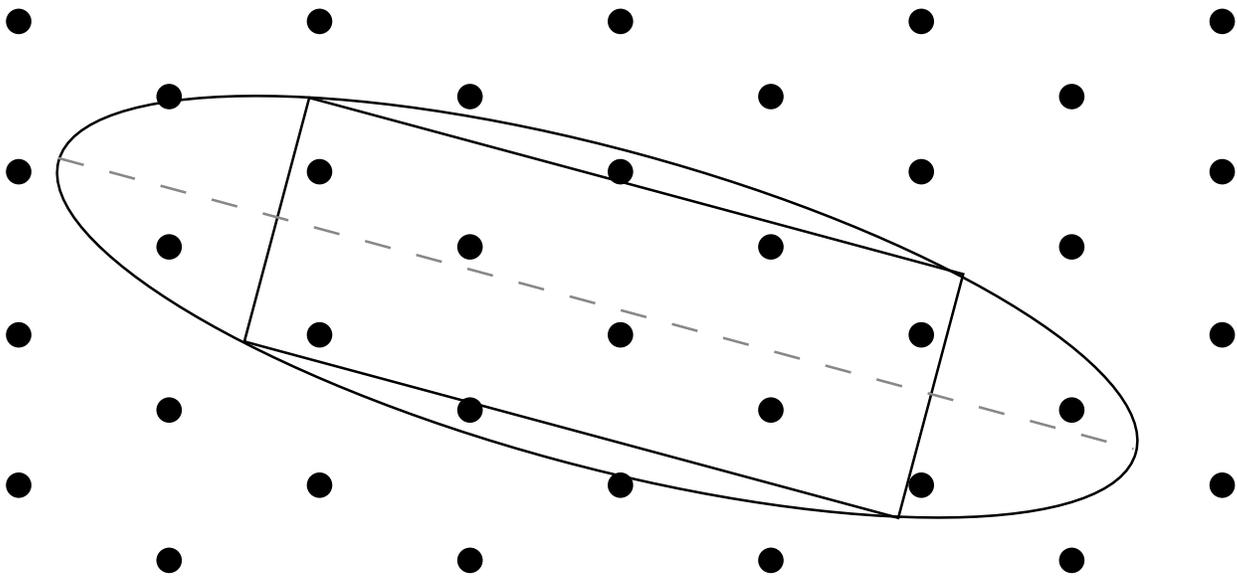
\noindent{Figure 3: Illustration of the intersection of the "active part" of
the shower (as described in the text) and the detector plane used in the
geometrical approach. The intersection, which is close to an ellipse, is
approximated by a rectangle as illustrated. The dots correspond to the detector
tanks.}

\vspace{1.cm}

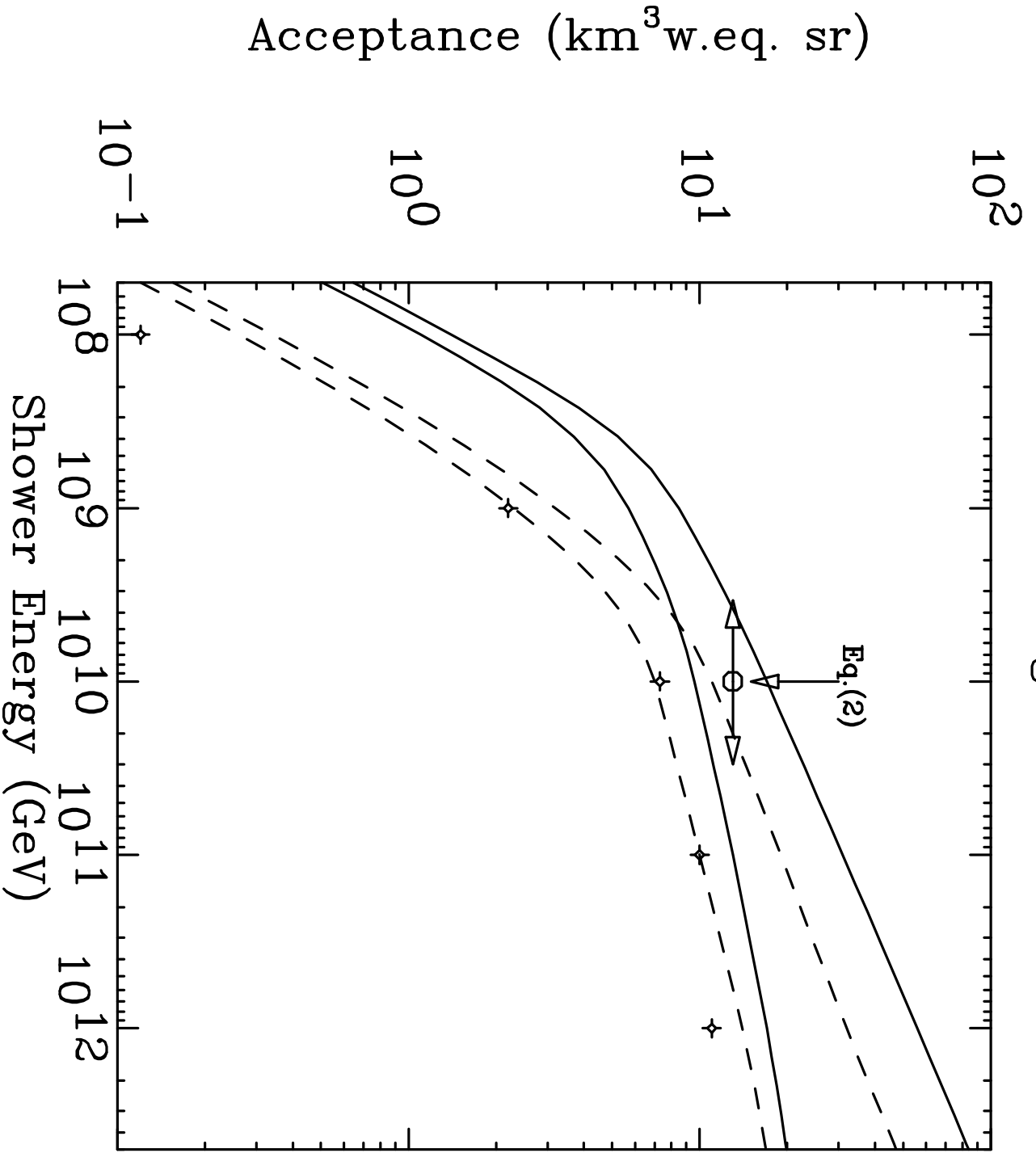
\noindent{Figure 4: Acceptance of the Pierre Auger detector to near horizontal
showers ($\theta_{zenith} \ge 75$ degrees). Volume units  are $km^3$ of water
equivalent. Crosses are for the results in the Monte Carlo approach, lines are
correspond to the geometrical integration for electromagnetic showers (dashed)
and hadronic (solid). The lower set of curves corresponds to only showers with
axis falling in the array and the upper set takes all showers into
consideration.}

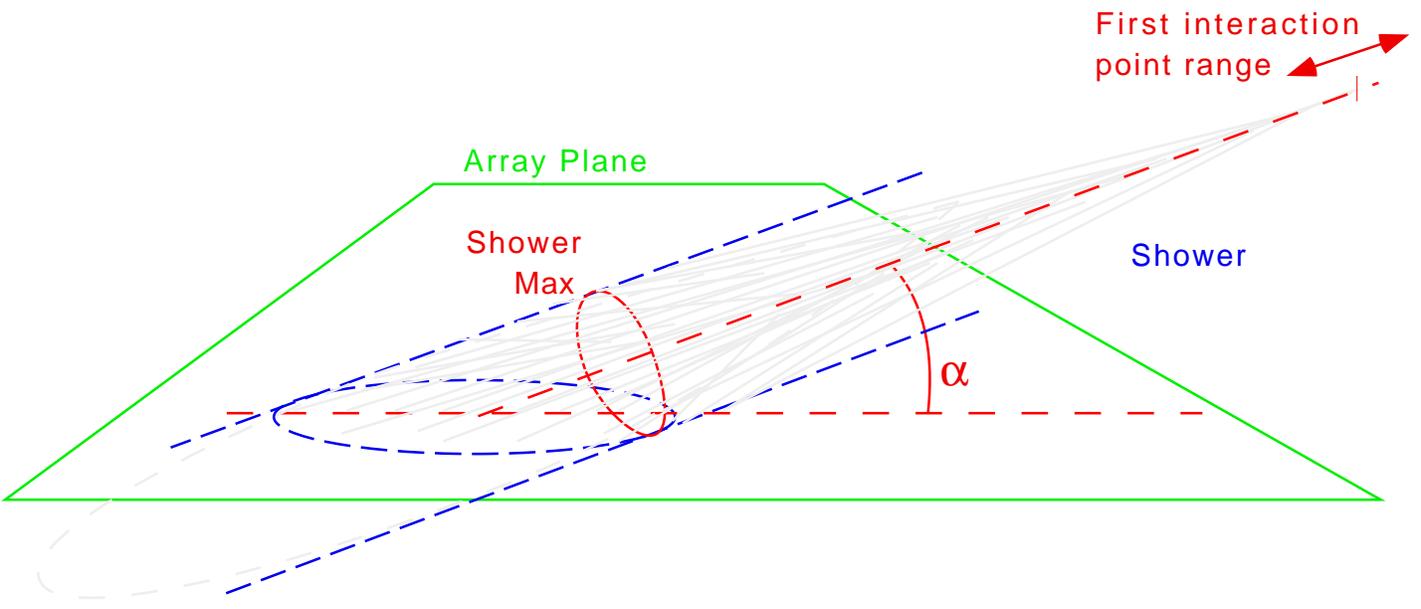
\end{document}